Title

# Kibble-Zurek mechanism of Ising domains

Kai Du[1], Xiaochen Fang[1], Choongjae Won[2], Chandan De[3,4], Fei-ting Huang[1], Fernando J. Gómez-Ruiz[5,6], Adolfo Del Campo[7,6], and Sang-Wook Cheong[1]*

1. Rutgers Center for Emergent Materials and Department of Physics and Astronomy, Rutgers University, Piscataway, New Jersey 08854, USA
2. Laboratory for Pohang Emergent Materials and Max Planck POSTECH Center for Complex Phase Materials, Department of Physics, Pohang University of Science and Technology, Pohang 37673, Korea
3. Center for Artificial Low Dimensional Electronic Systems, Institute for Basic Science (IBS), Pohang 37673, Korea
4. Laboratory of Pohang Emergent Materials, Pohang Accelerator Laboratory, Pohang 37673, Korea
5. Instituto de Física Fundamental IFF-CSIC, Calle Serrano 113b, Madrid 28006, Spain
6. Donostia International Physics Center, E-20018 San Sebastián, Spain
7. Department of Physics and Materials Science, University of Luxembourg, L-1511 Luxembourg, Luxembourg

* To whom the correspondence should be addressed. (E-mail: sangc@physics.rutgers.edu)

**Abstract**

**The formation of topological defects after a symmetry-breaking phase transition is an overarching phenomenon that encodes rich information about the underlying dynamics. Kibble–Zurek mechanism (KZM), which describes these nonequilibrium dynamics, predicts defect densities of these second-order phase transitions driven by thermal fluctuations. It has been verified as a successful model in a wide variety of physical systems, finding applications from structure formation in the early universe to condensed matter systems. However, whether topologically-trivial Ising domains, one of the most common and fundamental types of domains in condensed matter systems, also obey the KZM has never been investigated in the laboratory. We examined two different kinds of three-dimensional (3D) structural Ising domains: clockwise (CW)/counter-clockwise (CCW) ferro-rotation domains in $NiTiO_3$ and up/down polar domains in BiTeI. While the KZM slope of ferro-rotation domains in $NiTiO_3$ agrees well with the prediction of the 3D Ising model, the KZM slope of polar domains in BiTeI surprisingly far exceeds the theoretical limit, setting an exotic example where possible weak long-range dipolar interactions play a critical role in steepening the KZM slope of non-topological quantities. Our results**

**demonstrate the validity of KZM for Ising domains and reveal an enhancement of the power-law exponent and a possible reduction of the dynamic critical exponent *z* for transitions with long-range interactions.**

**Main text**

Phase transitions and their related phenomena lie at the core of modern statistical mechanics and condensed matter physics. At equilibrium, an intriguing aspect of second-order phase transitions is that systems with distinct order parameters can be described by the same set of static critical exponents, a hallmark of universality[1]. On the other hand, the dynamical aspect of phase transitions away from equilibrium, such as the formation of intriguing topological defects and domain structures, is much more elusive. The Kibble–Zurek mechanism (KZM) is a powerful theory that describes the universality of such nonequilibrium dynamics and predicts the density of topological defects after a continuous second-order phase transition driven at a finite cooling rate. The formation of topological defects in a scenario of spontaneous symmetry breaking was first proposed by Kibble in the early universe[2] and further analyzed by Zurek[3] paving the way to testing cosmological principles in the laboratory. KZM predicts that the density of topological defects $n_v$ should obey a power-law relation with the cooling rate $r$ characterized by a universal exponent $\beta_{\text{KZM}}$. Specifically, this universal exponent reads

$$\beta_{\text{KZM}}(z, \nu) = \frac{D\,\nu}{1 + z\nu} \qquad (1)$$

which only depends on the spatial dimension (*D*), the spatial critical exponent (*ν*), and the dynamical critical exponent (*z*) of the system. This universality essentially enables the testing of KZM in condensed matter systems inside the same universality class, which is impractical at the cosmic scale initially proposed.

Though possible, the search for ideal systems to test KZM remains challenging. To date, aspects of KZM have been experimentally examined in various systems such as superfluid $^4$He (ref. [4]), superconductors[5], liquid crystals[6], trapped ions[7,8], and Bose-Einstein condensates[9]. As a more recent prototype example, the density of structural Z6 topological vortices in rare-earth hexagonal manganites is found to match the prediction of the KZM very well with a universal KZM slope ($\beta_{\text{KZM}}$) of ~0.59 for various compounds[10]. Although it was initially proposed to simulate defects with nontrivial topological properties, KZM has also been applied to compute other non-topological quantities, such as excess energy[11], excess kinetic energy in cold gases [12], and even nonequilibrium coherence lengths with no reference to topological defects [13]. Actually, whenever an equation of motion can be written down for the order parameter (as in Ginzburg-landau models), KZM can be derived rigorously from finite-time scaling analysis[14]. However, fundamental three-dimensional (3D) Ising-type domains with two degenerate scalar order parameters have never been investigated in the context of KZM, despite their ubiquity in condensed matter systems. Whether these topologically-trivial binary domains still follows KZM and if their dynamics are different from those well-studied topological defects remain elusive. This is largely due to the lack of good material systems whose Ising domain populations can be easily imaged and compared in a wide range of cooling rates. Magnetic Ising domains are conventionally the most accessible Ising domains thanks to various developed magnetic imaging techniques. Unfortunately, the domain size in the ordered ferromagnetic phase is often dominated by the competition between material-specific parameters like exchange interactions, magnetic anisotropy, and demagnetization fields other than KZM, as exemplified in well-established magnetic domain theories[15]. By contrast, Ising domains can be easily frozen after a structural transition, making the latter an ideal testbed for intrinsic nonequilibrium dynamics described by KZM. In this work, we accomplish the testing of KZM by investigating two different 3D structural Ising domains: clockwise (CW)/counter-clockwise (CCW) ferro-rotational domains in $NiTiO_3$ and up/down polar domains in BiTeI. Exploiting our newly-developed approaches to domain imaging and analysis, we systematically explored in these materials the domain size dependence over a wide range of cooling rates. While Ising domains are found to follow the power-law relation predicted by KZM, we also discovered an exotic KZM slope steepening in BiTeI and provided its possible origin.

**KZM of 3D Ising ferro-rotational $NiTiO_3$**

$NiTiO_3$ crystallizes in the corundum structure (space group $R\bar{3}c$) at high temperatures and undergoes an order-disorder type structural transition to the ilmenite structure (space group $R\bar{3}$) at ~1297°C (ref. [16]), which is reported to be a second-order phase transition according to neutron studies[17,18]. At room temperature, the ordered stacking of $Ni^{2+}$ and $Ti^{4+}$ layers introduces an imbalance of oxygen rotation surrounding them and gives rise to a total net rotation in each unit cell. Depending on the stacking sequence, it can have two different ferro-rotational domains (Fig. 1a) that we refer to CW and CCW domains. This ferro-rotational order (also known as ferro-axial orders) has been recently recognized as a new type of ferroic order[19,20], adding an ideal 3D Ising-type system for the testing of KZM. However, the visualization of these ferro-rotational domains remains difficult as the inversion symmetry is not broken in ferro-rotational materials with space group $R\bar{3}$ (ref. [21]). It becomes even more challenging when it comes to the study of KZM, where huge areas and a large number of samples need to be characterized. So far, only transmission electron microscopy (TEM)[22,23] and microscope combined with the linear electrogyration effect[20] have shown the capability of visualizing ferro-rotational domain structures. Unfortunately, both techniques require delicate and excessive sample preparations, which make them impractical to deal with the extensive measurements required for probing the KZM.

To overcome this hurdle, we developed a new selective polishing technique that can easily reveal ferro-rotational domains in NiTiO$_3$. The circular motion of polishing coupled to the ferro-rotational domains creates a height difference between them, which can be captured by a circular differential interference contrast (CDIC) microscope[24] (Methods for details). Fig. 1b shows a typical CDIC image on the hexagonal *ab* plane of a NiTiO$_3$ crystal after the selective polishing, where curvy CW and CCW ferro-rotational domains are evident. The height difference between two domains is about 2 nm according to the atomic force microscopy image (Extended Data Fig. 1), indicating one of the ferro-rotational domains is polished more than the other after polishing in circular motions. Further dark-field TEM studies on quenched crystals confirm that these curvy domains are indeed ferro-rotational domains (Extended Data Fig. 2). The corresponding microscopic mechanism of selective polishing is beyond the scope of this work and needs to be further studied in the future. Nevertheless, the combination of selective polishing and optical observation is a new visualization tool, which enables convenient access to real-space ferro-rotational domain information with a wide range of fields of view.

Next, we try to obtain the density of defects (i.e., the domain density for Ising domains) in samples with different cooling rates based on experimental CDIC images. This is not straightforward as the size of these Ising domains is not well-defined and spans a very wide range due to its special topology. Therefore, a statistical analysis of a large number of domains is needed for fair comparisons of their domain sizes. Note that the conventional way of mapping domain-wall contours and counting the isolated domain area is not appropriate for Ising domains since the resulting domain density will become highly sensitive to local percolations other than the average granular domain size. To avoid that, we first manually converted the experimental CDIC image (Fig. 1b) to a black/white image (Fig. 1c) to facilitate further automatic analysis. Then, the black/white image is fed into a Matlab program to automatically calculate the number of domains in each horizontal and vertical pixel-line profile as exemplified in Fig. 1d and Fig. 1e, respectively. This statistical approach involving two orthogonal directions is crucial to eliminate statistical artifacts from possible domain elongations along a particular direction. Summarizing the total line profile length and dividing it by the counted total number of domains, we can obtain the average domain size *L* for the image. Finally, the density of defects $n_v$ can be derived by $n_v = L^{-2}$ (see Methods).

Utilizing the approach above, we systematically investigated the density of 3D Ising domains $n_v$ in NiTiO$_3$ as the function of cooling rate *r* across its ferro-rotational transition (Fig. 2a). Except for the quenched one, all other samples are annealed in a narrow temperature window (from 1310°C to 1260°C) with the labeled cooling rate and then quenched to room temperature to minimize the influence of possible coarsening (or phase ordering) effect[25]. Original CDIC images of each cooling rate and their corresponding processed black/white images for analysis are included in Extended Data Fig. 3. The linear fitting in the log-log plot between domain density $n_v$ and cooling rate *r* demonstrates a clear power-law relation between them. The fitted slope is the KZM exponent $\beta_{\text{KZM}}^{\text{Exp}} \approx 0.85$ for ferro-rotational domains in NiTiO$_3$, which is the first experimental value obtained in the 3D Ising universality class.

We next check its consistency against the theoretical prediction of KZM. The spatial critical exponent $v$ of 3D Ising universality class is known to be close to 0.63 according to numeric calculations[26–29]. On the other hand, the numerical values reported for the dynamical critical exponent $z$ do have some fluctuations depending on the numerical method[30–36], which have been summarized in Fig. 2b. According to equation (1), we estimate the theoretical/numerical KZM exponent $\beta_{\text{KZM}}^{\text{Num}}$ using a standard error propagation theory as:

$$\beta_{\text{KZM}}^{\text{Num}}(z, \Delta z; v, \Delta v) = \beta_{\text{KZM}}(z, v) \pm \sqrt{[\frac{\partial}{\partial z}\beta_{\text{KZM}}(z,v)]^2 \Delta z^2 + [\frac{\partial}{\partial v}\beta_{\text{KZM}}(z,v)]^2 \Delta v^2} \quad (2)$$

where $\Delta z$ and $\Delta v$ are the corresponding numerical uncertainty for the dynamic and spatial critical exponents, respectively. As a consequence, we obtain

$$\beta_{\text{KZM}}^{\text{Num}}(z, \Delta z; v, \Delta v) = \frac{D}{1+zv}[v \pm \frac{1}{1+zv}\sqrt{\Delta v^2 + v^4 \Delta z^2}] \quad (3)$$

The final calculated values of $\beta_{\text{KZM}}^{\text{Num}}$ are plotted in Fig. 2c with an average of $\overline{\beta_{\text{KZM}}^{\text{Num}}} \approx 0.81$. Our experimental result agrees well with this theoretical calculation within a small margin and clearly illustrates the validity of KZM for 3D Ising domains.

**KZM of 3D Ising polar BiTeI**

Having established the validity of KZM for Ising ferro-rotational domains, we now extend our investigation to the realm of Ising polar domains. Among popular van der Waals materials nowadays, BiTeI is a layered semiconductor hosting Ising polar domains with a large bulk Rashba effect[37] and chiral electronic excitations[38]. Different stacking sequences along the hexagonal $c$ axis with either Te or I terminations give rise to its up/down polar structures (Fig. 3a) at room temperature (space group *P*3*m*1). In the high-temperature range, it was reported to have a structural transition at ~470°C indicated by a sudden resistivity change below its melting point at ~560°C (ref. [39]). However, the exact nature of this transition is still unclear due to its instability at high temperatures. Typical X-ray or neutron studies at high temperatures are technically challenging and still lacking so far. Here, we experimentally demonstrate that this structural transition is the polar-nonpolar transition through which its polar domain size can be systematically tuned to test the KZM.

Single-crystalline BiTeI was synthesized by a modified Bridgman method (see Methods). The millimeter-sized flat surface can be easily exfoliated, see Fig. 3b, demonstrating the high quality of these crystals. Regarding the observation of polar domains in BiTeI, a similar difficulty also applies since existing imaging methods like scanning tunneling microscopy (STM)[40] and TEM can not handle imaging tasks requiring a large field of view. Alternatively, we surprisingly found that these polar domains can be easily probed by piezoresponse force microscopy (PFM) shown in Fig. 3c, even though these BiTeI crystals are actually metallic. In principle, theoretically-allowed piezoresponse in BiTeI (ref. [41]) should have been screened by nonstoichiometry-induced charge carriers[37,42]. Although it is out of the scope of the current work, the intriguing mechanism behind this paradoxical fact merits further investigations in the future. Nevertheless, we managed to map out polar domains of BiTeI by PFM and convert the PFM image to a black/white image (Fig. 3d) for further domain size analysis (Fig. 3e). Following the same analytical procedures introduced for NiTiO$_3$, one can now study the cooling rate dependence of polar domain sizes straightforwardly. A series of BiTeI crystals were prepared with different cooling rates from 500°C to 400°C, and then 100 K/h-cooled from 400°C to room temperature. Their polar domain sizes do vary consistently with corresponding cooling rates (Extended Data Fig. 4), which confirms the aforementioned transition at ~470°C as the polar structural transition. Since the ionic size of Tellurium and Iodine are closely similar, the most plausible high-temperature structure is expected to be a Tellurium- and Iodine-disordered non-polar phase due to possible intermixing above this transition.

Although BiTeI has a van der Waals-bonded structure and a weak coupling along the *c* axis, whether these polar domains can be considered as a quasi-2D Ising order or still a 3D Ising order remains unknown. Utilizing dark-field TEM, we investigated and compared polar domains on both the *ab* plane and the side surface of BiTeI. Surprisingly, these polar domains have identical shapes and sizes on both surfaces without any directional preference (Figure 4a and 4b), which manifests their 3D nature. Therefore, the polar domains in BiTeI should belong to the same 3D Ising universality class as the ferro-rotational domains in NiTiO$_3$. Plotted in the log-log scale, Fig. 4c illustrates again a clear power-law relation between domain density $n_v$ and cooling rate *r*, where we obtain a KZM exponent $\beta_{\text{KZM}}^{\text{Exp}} \approx 1.1$ for 3D polar domains in BiTeI. Note that KZM with the power-law scaling relies on a continuous phase transition and in particular, on the existence of critical slowing down near the second-order transition. Therefore, our results imply that the transition at ~470°C in BiTeI is likely a second-order polar transition, above which iodine and tellurium orderings disappear. Surprisingly, this KZM exponent is much larger than the theoretical value ($\overline{\beta_{\text{KZM}}^{\text{Num}}} \approx 0.81$), which suggests additional factors may be involved in the dynamics of polar domains in BiTeI and contributed to the exotic steepening of the KZM slope. We shall discuss its possible origin next.

## Discussions

First of all, we would like to show that the KZM slope steepening in BiTeI is due to its intrinsic characteristic other than extrinsic artifacts. We have reexamined one 0.5 K/h-cooled and one 20 K/h-cooled BiTeI crystal with an additional coarsening at 400°C for 300 hours. No significant changes in domain density are found after excessive coarsening (green and blue data points in Extended Data Fig. 5), which rules it out as the origin of the steepened KZM slope. Moreover, a previously slowly-cooled (0.5 K/h) crystal with a low domain density can be converted freely to the one with a high domain density by a post-annealing (40 K/h, pink data point in Extended Data Fig. 5), which illustrates the excellent uniformity and the high quality of our crystals. This clearly shows the steepened KZM slope does not originate from trivial pinned chemical defects in different BiTeI crystals but an intrinsic property of its polar transition.

Among the three intrinsic parameters that determine the KZM exponent in equation (1), the 3D nature of polar domains in BiTeI has been clearly established based on our TEM results. Assuming the spatial critical exponent ($v \approx 0.63$) of the 3D Ising model, an extraordinary dynamical critical exponent ($z$) in BiTeI is left to be the most plausible cause of its steep KZM slope. Taking the experimental KZM slope value, we obtain $z \approx 1.14$ for BiTeI, which is exceptionally small compared to the typical ballpark of $z$ around ~2.12 (Fig. 2b) for the 3D Ising model. This means the relaxation time ($\tau$) near the polar transition in BiTeI has a much broader and slower response to the reduced temperature $T_{red}$ according to the relation $\tau \sim T_{red}^{-vz}$ (Extended Data Fig. 6). To figure out the mechanism behind it, we can gain useful insights by comparing the difference between the case of $NiTiO_3$ and BiTeI. While both of them host 3D Ising order with an order-disorder phase transition at high temperatures, distinctive ferro-rotational order and polar order make the major difference between the two compounds. The ferro-rotational order in $NiTiO_3$ is determined by pure local structural interactions at the transition. Therefore, it is supposed to fit the tenets of KZM which considers only local interactions to justify a well-defined causal front, dictated by the speed of second sound, giving rise to independent choices of the broken symmetry in spatially separated regions of the system. On the other hand, the polar order in BiTeI can potentially have additional long-range dipolar interactions which may induce the deviation from the ideal KZM prediction. These long-range dipolar interactions can be weak due to the screening effect from mobile carriers in BiTeI at room temperature. However, they can still interfere with the domain dynamics near the high-temperature transition when polar orders are not frozen and the screening is supposed to be reduced due to increased resistivity. Therefore, weak long-range dipolar interactions are most likely responsible for the increased KZM slope and exceptionally small $z$ in BiTeI. Theoretically, the effect of adding long-range interactions to the transition on the spatial critical exponent ($v$), dynamical critical exponent ($z$), and the KZM is still an open problem for 3D systems due to the limited computational resources at present. For 1D like Ising systems, limited theoretical studies so far often report contradicting results[43,44].

Additionally, recent studies for the ferromagnetic long-range Ising model showed that all the critical exponents can significantly change according to the power-law exponent for the long-range interaction[45]. Therefore, our current KZM study cannot fully rule out possible spatial critical exponent (*v*) deviations in the presence of long-range interactions. To experimentally disentangle the contributions from *v* and *z* to the KZM slope requires accessing the relaxation freeze-out time[3] near the transition, which follows an independent power-law with the driving rate governed by an exponent that depends only on the product of *v* and *z*. Unfortunately, this is not feasible in systems like BiTeI and $NiTiO_3$ due to their high transition temperatures. Nevertheless, our experimental results clearly support the scenario with an increased KZM slope in the presence of weak long-range interactions[43].

Beside the comparison between $NiTiO_3$ and BiTeI within 3D Ising university class, one can also gain more insights by comparing BiTeI and hexagonal manganites, both of which have possible dipolar interactions from their polar orders. In contrast to Ising polar domains, $Z_6$ topological vortices in rare-earth hexagonal manganites exhibit an excellent consistency with the theoretically-predicted KZM slope[10] despite the presence of dipolar interactions. An intriguing implication of these experimental facts is that topologically-protected $Z_6$ vortices in hexagonal manganites are almost immune to dipolar interactions and KZM is dominating. On the other hand, KZM prediction for non-topological Ising domains will be vulnerable to possible long-range interactions (e.g. dipolar interactions in BiTeI), although it can still work well for Ising domains with only local short-range interactions (e.g. ferro-rotational domains in $NiTiO_3$).

**Summary**

To conclude, we have developed innovative imaging techniques and analytical methods for the investigation of the ferro-rotational transition in $NiTiO_3$ and the polar transition in BiTeI. With these new tools, we managed to validate that fundamental Ising-type domains in condensed matter systems also follow the general prediction of KZM. In addition, we reveal that the dynamical critical exponent *z* of these non-topological Ising domains can be vulnerable to other factors and can be dramatically reduced with the presence of weak long-range interactions, leading to an increased KZM exponent. Our findings open the realm of exploring ubiquitous Ising domains in terms of intriguing KZM in 3D Ising universality class and demonstrate the important role of potential long-range interactions in determining KZM exponent in non-topological systems.

**Methods**

**Crystal growth and selective polishing of $NiTiO_3$**

High-quality single crystals of $NiTiO_3$ were prepared by a floating zone technique. The $NiTiO_3$ powder is prepared by a solid-state reaction of NiO and $TiO_2$ at 1350°C with several intermediate grindings. The single-phase $NiTiO_3$ powder was filled in a rubber tube and pressed into a rod by a hydrostatic pressure of 110 MPa. Then, the rod was heated at 1350°C for 24 hours. The single crystal was grown in an optical floating zone furnace at 3mm/hour under $O_2$ flow.

The hexagonal *ab* plane was orientated by Laue and mechanically polished on diamond lapping films with circular motions. The crystal surface was finally finished by vibratory polishing in colloidal silica slurry for 30 minutes (particle size: ~20 nm on average) with the same circular motions to reveal ferro-rotational domains. Optical images of ferro-rotational domains were taken by a Zeiss microscope in the CDIC mode. The topography of the polished surface was then further examined by atomic force microscopy.

**Automatic analysis of Ising-type domains**

For $NiTiO_3$, CDIC images of ferro-rotational domains are manually converted to black/white images to eliminate artifacts from defective areas. Then, each horizontal and vertical lineprofile of the image was extracted to calculate the total number of domains and the total lineprofile length by homemade Matlab codes. The average domain size *L* for the image can be obtained by dividing the total lineprofile length by the total number of domains. For BiTeI, homemade Matlab codes were used to first convert PFM images into black/white images by choosing the median contrast level as the threshold. Then, the same Matlab program is used to calculate the average domain size *L* for each image.

**Crystal growth and PFM imaging of BiTeI**

High-quality single crystals of BiTeI were prepared by a modified Bridgman technique. Stoichiometric bismuth shot, tellurium powder, and iodine were sealed in an evacuated quartz tube with a little excess of Iodine. The sealed ampules were annealed at 700°C for 24 hours with a heating rate of 20~30°C/h. After a furnace cooling, the ampule is flipped and placed upside down in a vertical furnace, and heated at 700°C for 12 hours. Then the ampule is cooled down to 400°C for a different cooling rate with a small temperature gradient. For post-annealings, crystals were placed in the evacuated quartz tube and heated only up to 500°C. Freshly-cleaved hexagonal *ab* surface of BiTeI crystal was scanned by PFM with a conductive Pt-coated contact-mode tip at room temperature. AC 1 V was applied to the tip at 45 kHz during the scan.

**Transmission electron microscopy (TEM)**
$NiTiO_3$ and side-view BiTeI specimens were prepared by mechanical polishing followed by Ar-ion milling. Thin BiTeI specimens with large *ab* plane surfaces were prepared by mechanical exfoliation followed by Ar-ion milling at the liquid nitrogen temperature. All specimens were studied using a JEOL-2010F field-emission TEM at room temperature.

## Data availability

The data that support the findings of this study are available from the corresponding author on reasonable request.

## Acknowledgements

We thank Dr. Shizeng Lin and Prof. Wojciech H. Zurek for their helpful comments on this work. The work at Rutgers University was supported by the DOE under Grant No. DOE: DE-FG02-07ER46382. The work at Pohang University was supported by the National Research Foundation of Korea funded by the Ministry of Science and ICT (grant No. 2020M3H4A2084417). FJG-R acknowledges financial support from European Commission FET-Open project AVaQus GA 899561.

## Competing interests
The authors declare no competing interests.

## Contributions
S.C. Initiated and guided the project; C.W. and C.D. grew the crystals; K.D. & X. F. prepared and measured samples; K.D. did data analysis; F.H. did TEM measurements; F.J.G. and A.D.C. performed theoretical analysis; K.D., F.J.G., A.D.C., and S.C. wrote the paper.

**Figure legends**

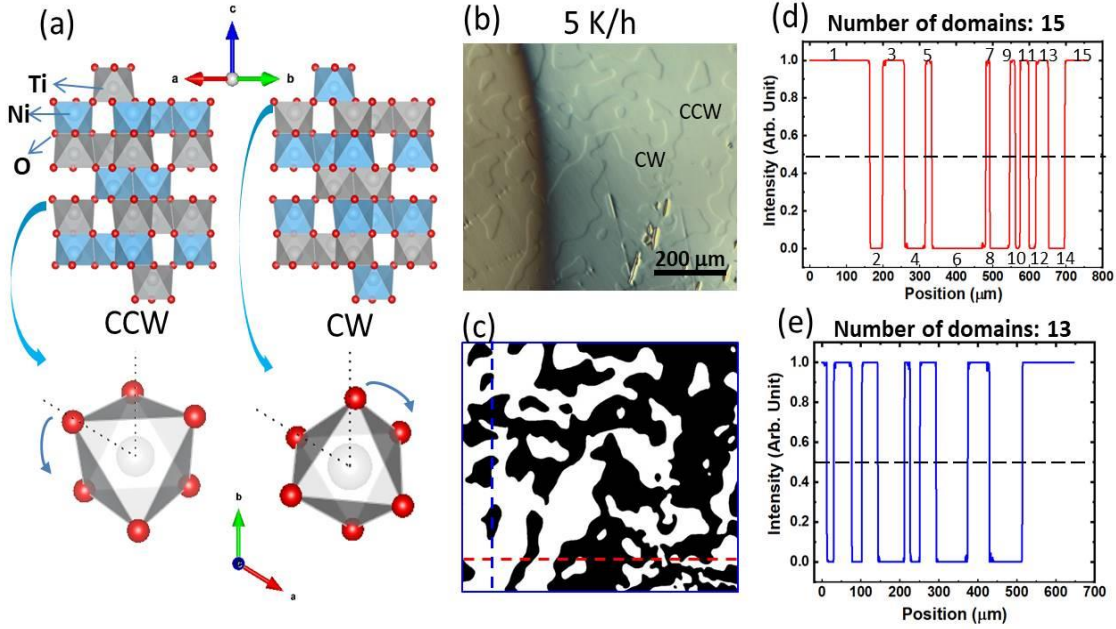

**Figure 1: The crystal structure and ferro-rotational domains of NiTiO$_3$. (a)** Counter-clockwise (CCW) and clockwise (CW) ferro-rotational structures of NiTiO$_3$ at room temperature. The lower parts are top views of oxygen cages with the corresponding sense of rotation. **(b)** CDIC optical images of the hexagonal *ab* plane of a 5 K/h-cooled NiTiO$_3$ after selective polishing. Clear ferro-rotational domains are observed. **(c)** The corresponding converted black/white image of (b). **(d-e)** Line profiles of the horizontal red dashed and the vertical blue dashed line in (c).

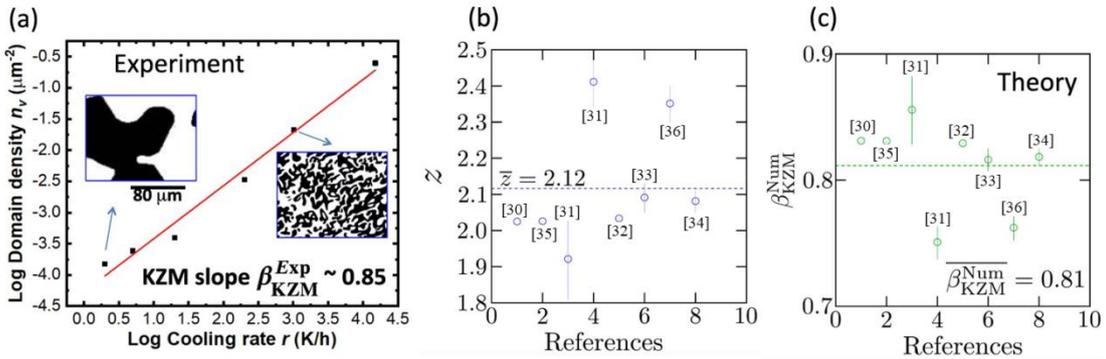

**Figure 2: KZM of 3D Ising ferro-rotational domains in NiTiO$_3$. (a)** The ferro-rotational domain density as a function of cooling rate in the log-log plot, with a fitted KZM slope of ~0.85. Insets are selected ferro-rotation domain images on the same scale. **(b)** Summary plot of numerically computed dynamical critical exponent z for the 3D Ising model from the literature. The marker plot includes a bracketed number.

These enumerations correspond with the number in the list of references. **(c)** Theoretically computed KZM slopes based on $z$ values provided in (b) with an average KZM slope of ~0.81, which agrees well with the experimental result in (a). For 3D Ising universality class, theoretically computed spatial critical exponent $v \approx 0.63$ is used for the calculation[26–29].

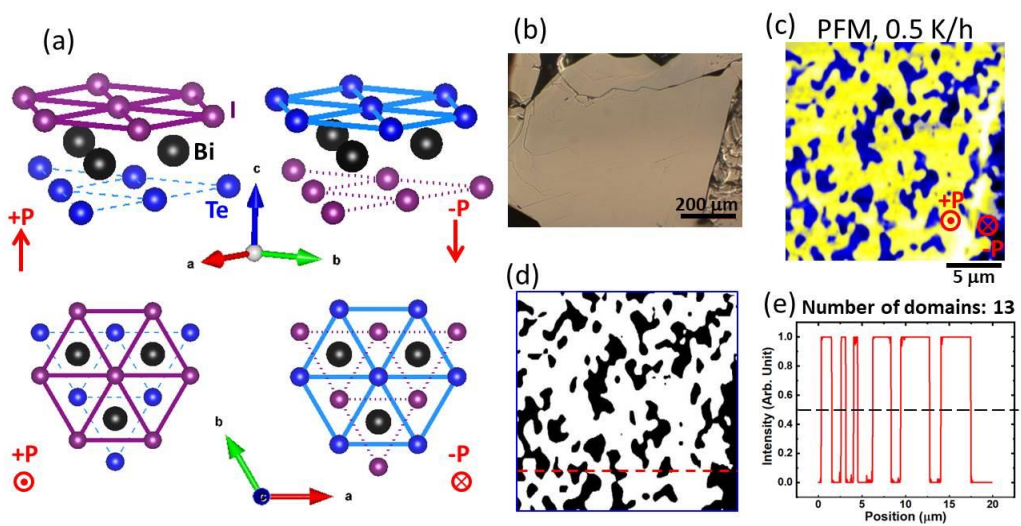

**Figure 3: The crystal structure and polar domains of BiTeI. (a)** Polar structures of BiTeI at room temperature. The lower parts are top views of different polar domains with either an I or Te termination. **(b)** Optical image of a freshly-cleaved hexagonal *ab* surface of BiTeI. **(c)** PFM image of a 0.5 K/h-cooled BiTeI crystal, showing Ising-type polar domains. **(d)** The corresponding converted black/white image of (c). **(e)** The line profile of the horizontal red dashed line in (d).

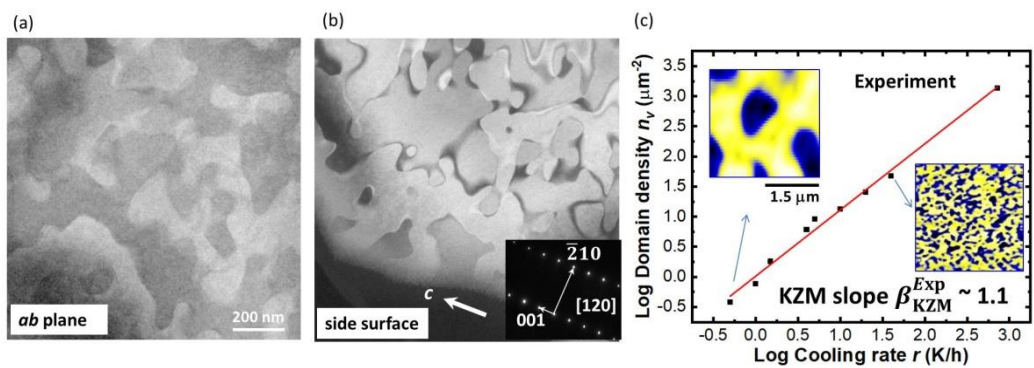

**Figure 4: KZM of 3D Ising polar domains in BiTeI. (a)** Dark-field TEM image of polar domains on the *ab* plane. **(b)** Side-view dark-field TEM images of polar domains reveal their 3D nature. Insets are corresponding selected area electron diffraction (SAED) pattern. **(c)** The polar domain density as a function of cooling rate in the log-log plot, with a fitted KZM slope of ~1.1. Insets are selected polar domain images on the same scale.

# Extended data

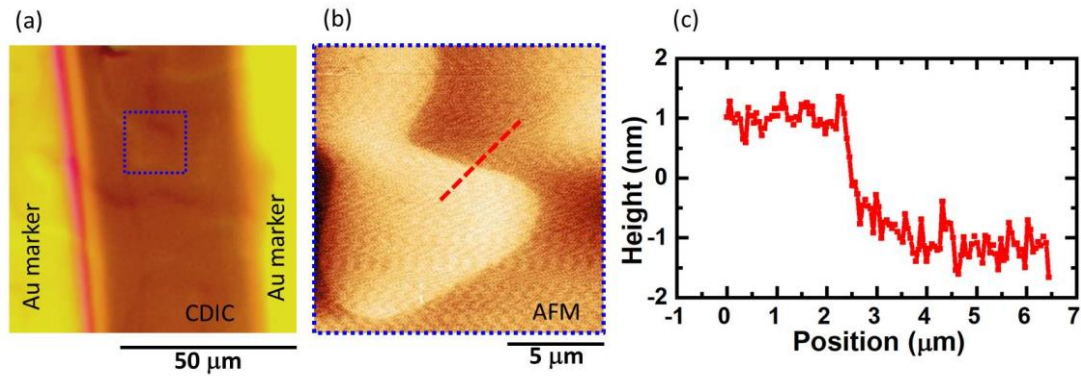

**Extended Data Fig. 1: Ferro-rotational domains after polishing in NiTiO$_3$.** **(a)** CDIC image of NiTiO$_3$ surface after polishing. **(b)** Corresponding atomic force microscopy topography image of the blue dotted region in (a). **(c)** Corresponding line profile of the red dashed line in (b).

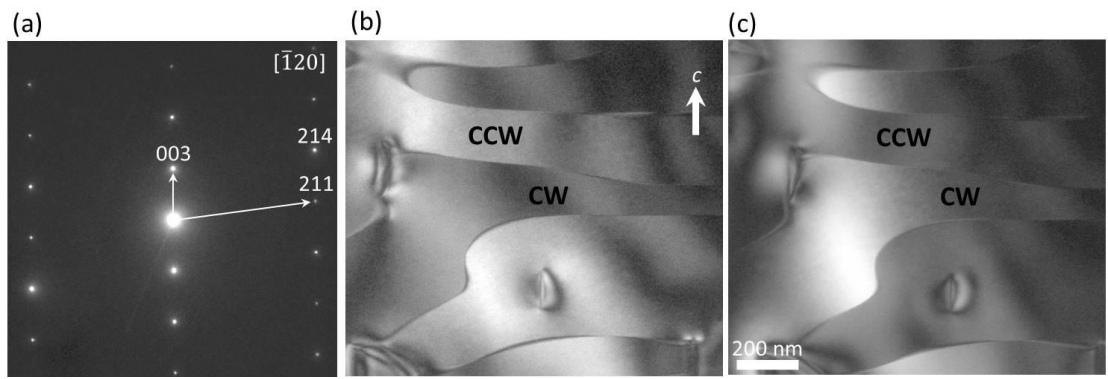

**Extended Data Fig. 2: Confirming ferro-rotational domains in NiTiO$_3$ by TEM.** **(a)** Selected area electron diffraction (SAED) pattern of quenched NiTiO$_3$ single crystal along [$\bar{1}$20]. Side-view dark-field TEM images by selecting spots **(b)** (214) and **(c)** (211), showing CCW and CW ferro-rotational domains. The zone [$\bar{1}$20] is indexed in respect to CCW ferro-rotational domain and become zone [$\bar{2}$10] in respect to CW domain. Note that the Bragg spot (214) is converted into (12$\bar{4}$) in the reciprocal space by the twofold rotation as two ferro-rotation domains do in real space. This produces a corresponding contrast difference between the two ferro-rotational domains that allows us to uniquely confirm the existence of ferro-rotational domains.

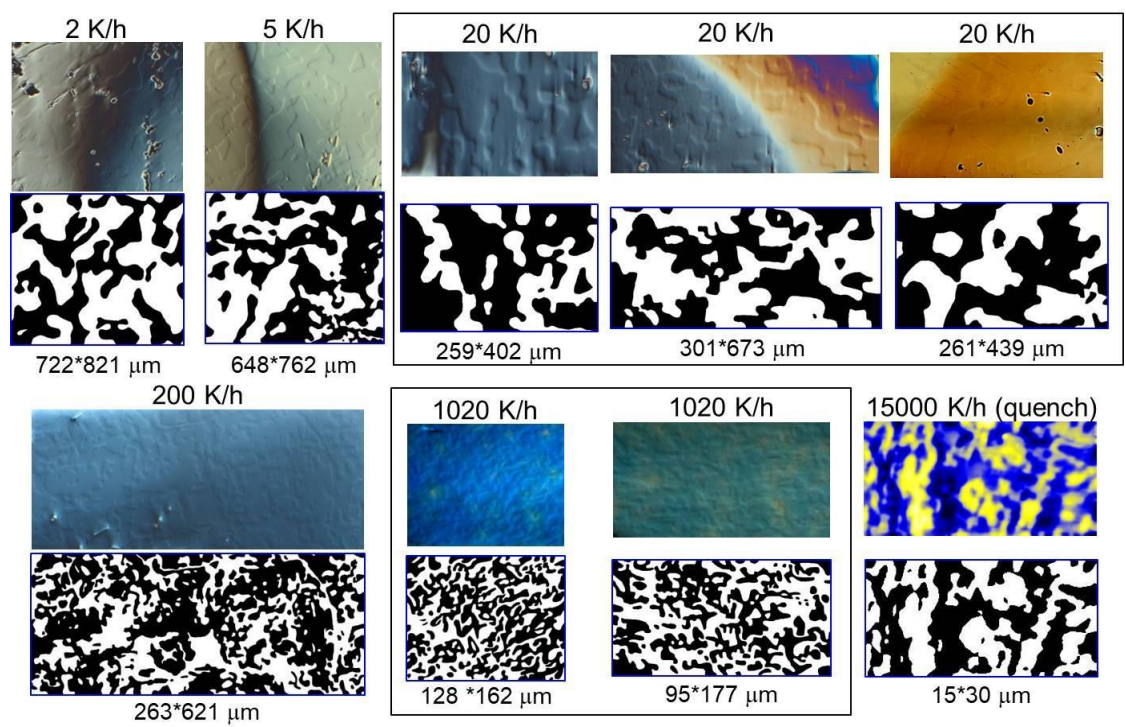

**Extended Data Fig. 3: CDIC images and corresponding black/white images of ferro-rotation domains for KZM plot in NiTiO$_3$ crystals.** Their image sizes and cooling rates are labeled individually. Images of crystals with the same cooling rate are grouped by black squares, and their average domain density is used for the KZM plot in Fig. 2(a). The cooling rate of the quenched crystal was estimated to be 15000 K/h at its ferro-rotational transition temperature. Due to its small domain size, the domain pattern of the quenched sample was imaged by AFM topography instead of the typical CDIC optical microscope method.

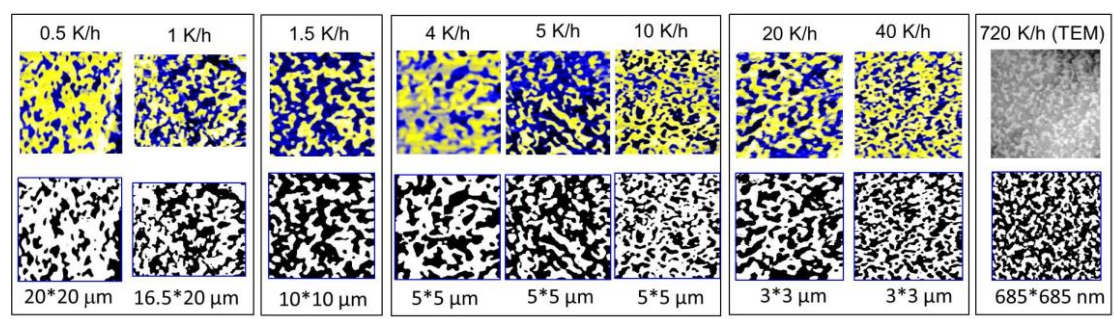

**Extended Data Fig. 4: PFM/TEM images and corresponding black/white images of polar domains for KZM plot in BiTeI.** Their image sizes and cooling rates are labeled individually. Images with the same scale are grouped by black squares to facilitate an easier comparison.

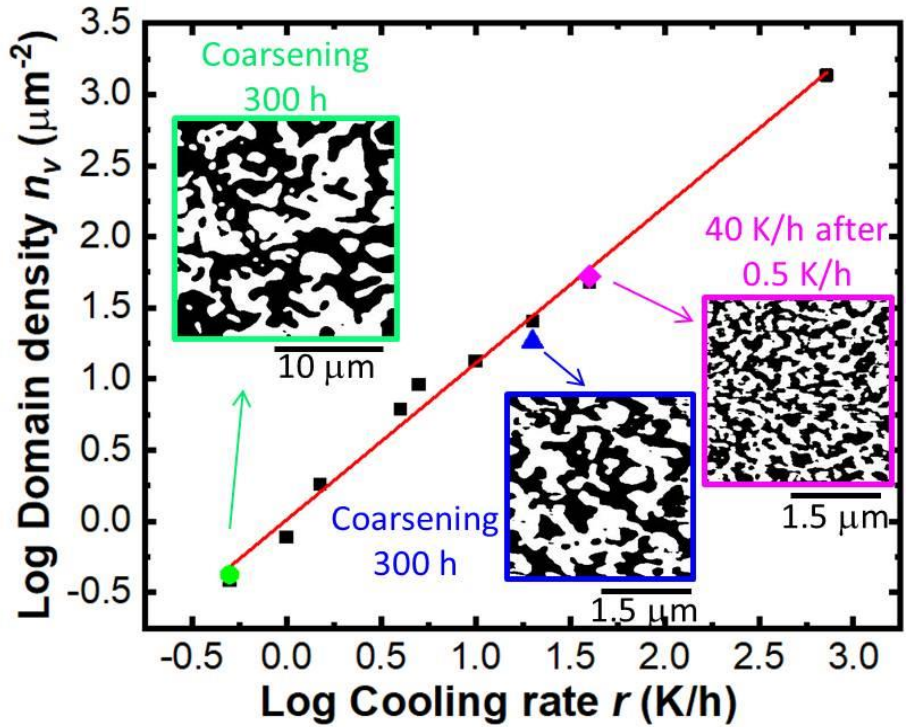

**Extended Data Fig. 5: Ruling out artifacts from the coarsening effect and chemical defects.** One 0.5 K/h-cooled (green data point) and one 20 K/h-cooled (blue data point) BiTeI crystal with an additional coarsening at 400°C for 300 hours are plotted with regularly-collected KZM data. No significant changes in domain density are found after excessive coarsening. An additional post-annealing (40 K/h, pink data point) can convert a previously slowly-cooled (0.5 K/h) crystal to a state with a high domain density, which illustrates polar domains are not pinned by chemical defects.

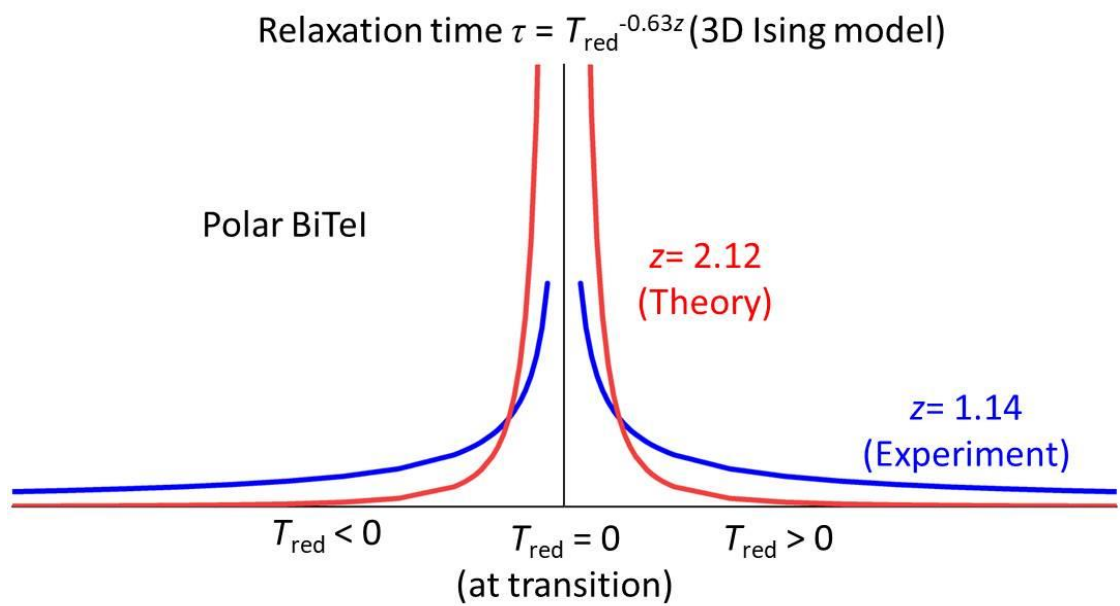

**Extended Data Fig. 6: Relaxation time as a function of reduced temperature with possible small dynamical critical exponent $z$ in BiTeI.** For 3D Ising model, the spatial critical exponent $v \approx 0.63$. The surprisingly small dynamical critical exponent $z$ from our experiments indicates that the relaxation time ($\tau$) near the polar transition in BiTeI has a much broader and slower response to the reduced temperature $T_{\text{red}} = (T-T_c)/T_c$, where $T_c$ is the polar transition temperature.